\documentclass[preprint,aps,prc]{revtex4}

\usepackage{epsfig}

\begin{document}

\title{Estimation of charm particles elementary cross sections relevant to GSI-CBM project.}
\author{Egle Tomasi-Gustafsson}
\email{etomasi@cea.fr}
\affiliation{\it DAPNIA/SPhN, CEA/Saclay, 91191 Gif-sur-Yvette Cedex,
France}
\author{Michail P. Rekalo }
\affiliation{\it National Science Center KFTI, 310108 Kharkov, Ukraine}
\begin{abstract}
This note collects formulas relative to the energy dependence of the cross sections for open charm and $J/\psi$ production for $NN$-collisions at threshold. It is a basis for the best input to MonteCarlo calculations, for associative charm particle production in nucleon-nucleon, nucleon-nucleus, ion-ion and proton-antiproton collisions.
\end{abstract}
\maketitle
\section{The reaction $p+p\to\Lambda_c^+ +\overline{D^0} +p$ }
\begin{figure}
\mbox{\epsfysize=5.cm\leavevmode \epsffile{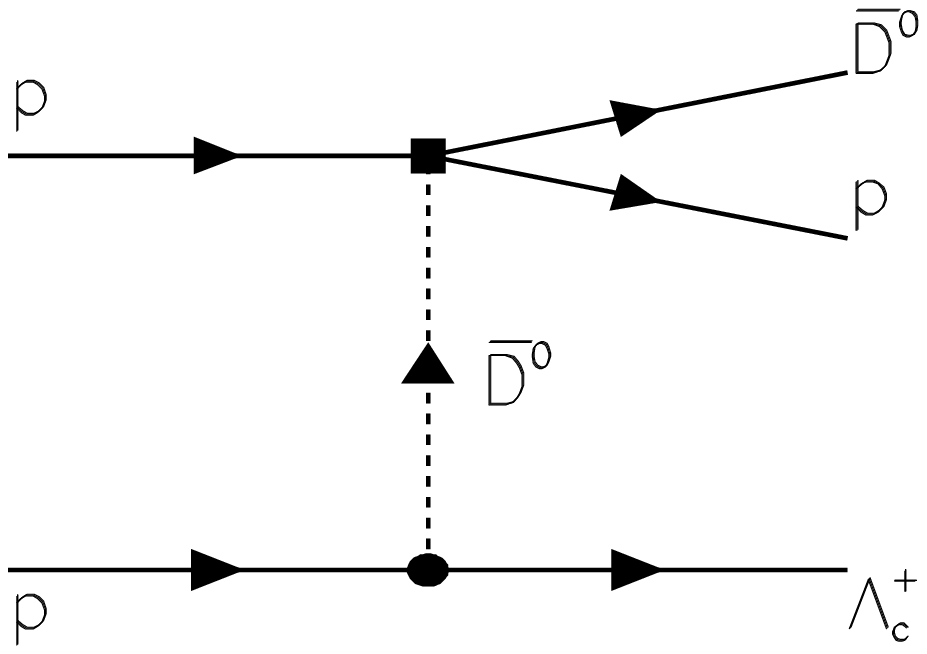}}
\vspace*{.2 truecm}
\caption{Feynman diagram for $p+p\to\Lambda_c^+ +\overline{D}^0 +p$. 
}
\label{fig:fig1}
\end{figure}

The threshold behavior of the total cross section for the reaction $p+p\to\Lambda_c^+ +\overline{D^0} +p$ can be described by:
\begin{equation}
\sigma(pp\to\Lambda_c^+ \overline{D^0}p)\simeq 0.2(Q/0.1 ~\mbox{GeV})^2\mbox{ nb}.
\label{eq:eq15}
\end{equation}
with 
\begin{equation} 
Q=\sqrt{s}-(m+M+\mu),
\label{eq:st2}
\end{equation}
$m$ is the nucleon mass, $M(\mu)$ is the mass of $\Lambda_c (D)$. This  estimation is based 
on $D-$exchange mechanism, Fig. \ref{fig:fig1}, and on the comparison with the processes $p+p\to\Lambda +K^+ +p$ ($K-$exchange), assuming $SU(4)$ symmetry\cite{Re01}. 
The threshold $Q^2$-behavior ( $Q$ is the energy excess over threshold) ,is the phase-space for a three-body reaction \cite{Mo02}.

\begin{figure}
\mbox{\epsfysize=15.cm\leavevmode \epsffile{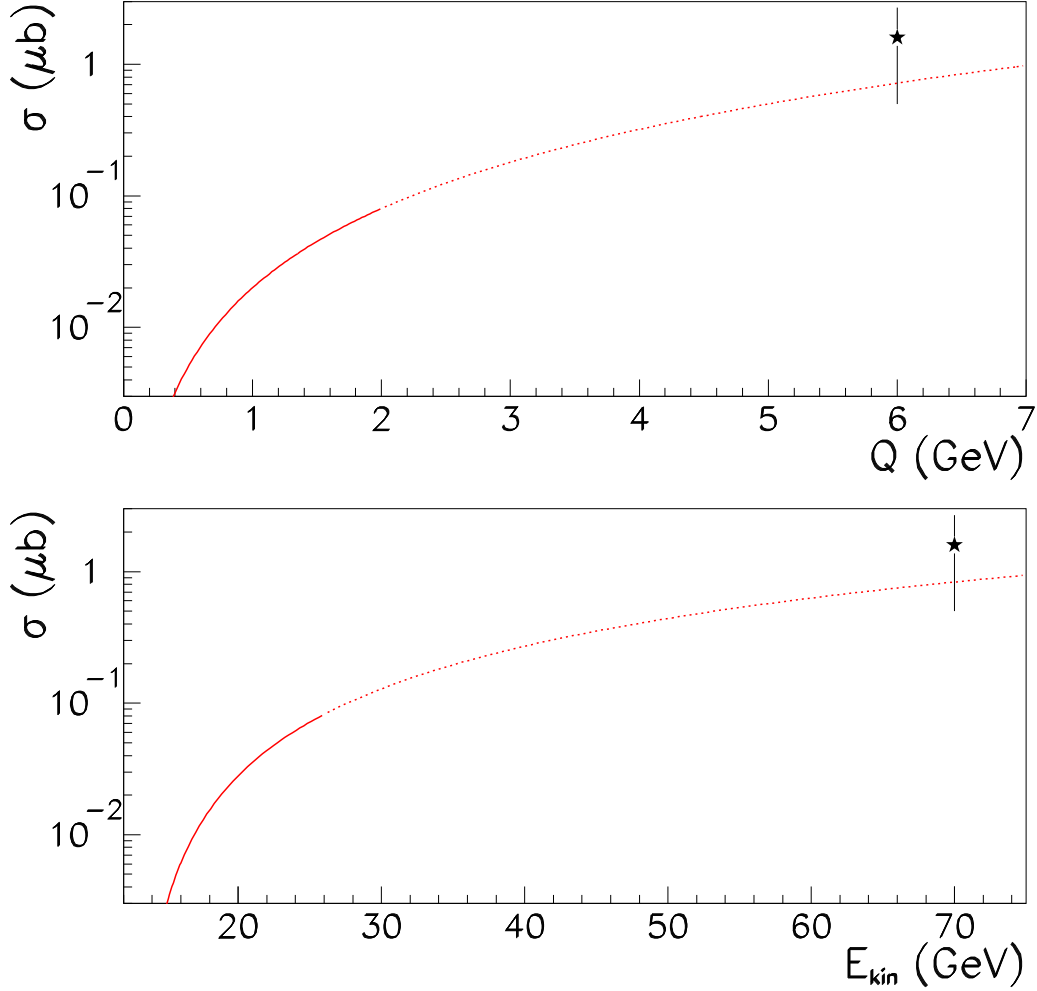}}
\vspace*{.2 truecm}
\caption{Cross section for the reaction $p+p\to\Lambda_c^+ +\overline{D^0} +p$ as a function of the energy excess over threshold, $Q$, (top) and of the proton kinetic energy (bottom)
}
\label{fig:fig2}
\end{figure}

In Fig. \ref{fig:fig2} we report the threshold behavior of the cross section for the reaction $p+p\to\Lambda_c^+ +\overline{D^0} +p$ as a function of $Q$, (top) and of the proton kinetic energy (bottom).

No experimental data exist in the threshold region. The lowest energy where the open charm cross section has been measured, $E_p$=70 GeV, $\sqrt{s}=11.46$ GeV, corresponds to $Q=6$ GeV. From Eq.
(\ref{eq:eq15}) we find, for this energy: $\sigma(pp\to\Lambda_c^+ \overline{D^0}p)\simeq 0.7$ $\mu$b, to be compared to the experimental value $\sigma(pp\to~charm )=1.6^{+1.1}_{-0.7}~ (stat)\pm 0.3~ (syst)$ $\mu$b \cite{Am01}. The calculation rigorously hold only for $Q< \mu$.

\section{Total cross section for open charm production}

The \underline{total cross section for $pp$-interaction} near threshold can be approximated as: $$\sigma_{tot}(pp)\simeq 40~ {\mbox~mb} {\mbox~at} ~\sqrt(s)\ge 10 {\mbox~GeV} $$ and it flattens at larger energies.

The \underline{semi-inclusive cross section for charm particle production} has been calculated in \cite{Ga02}, in analogy with strange particle production, and it has been found
$$\sigma(pp\to X\Lambda_c)\simeq 40{\mbox~\mu b}$$
depending on the choice of the cut off parameter for the form factor, which has to be introduced in the vertex of the diagram. Let us note that this number is larger than the direct experimental value \cite{Am01}. Using these cross section, one can find:
$$R_D=\displaystyle\frac{\sigma(pp\to \Lambda_c X)}{\sigma_{tot}(pp)}=(0.25\div 1)\cdot 10^{-3}$$

Following \cite{Ga02}, in analogy with strange particle production in $\gamma N$-collisions, in principle one should take into account another contribution, $D^*$-exchange,  which may increase the cross section of a factor four. Therefore:
$$R_{D+D^*}\simeq (1\div 4)\cdot 10^{-3}$$
which is in agreement with the experimental value  of 40$\div$ 200 $\mu$b at $\sqrt{s}$=62 GeV \cite{Ba91}.
\subsection{Isotopic relations }
The models \cite{Re01,Ga02} allow to calculate the cross section for all channels of exclusive charm particle production, not only for $pp$- but also for $np$-collisions. The starting point of such calculations is the isotopic structure of the matrix element ${\cal M}$ for $NN\to Y_c+\overline{D}+N$, which is model independent.

We give here the necessary relations based on the 
 isotopic invariance of the strong interaction,  with  $I(\Sigma_c)=1$, $I(\Lambda_c)=0$, $I(\overline{D})=1/2$:
\begin{equation}
\begin{array}{llrrrr}
1)&{\cal M}(pp\to \Sigma_c^{++}D^- p)=
&A_{11}&+\sqrt{2}A_{10}, &  & \nonumber \\
2)&{\cal M}(pp\to \Sigma_c^{++}\overline{D^0} n)=
&A_{11}&-\sqrt{2}A_{10},&  &\nonumber \\
3)&{\cal M}(pp\to \Sigma_c^+\overline{D^0} p)=
&-\sqrt{2}A_{11}, & & &\label{eq:amp} \\
4)&{\cal M}(np\to \Sigma_c^{++}D^- n )=
&~A_{11}& &+\sqrt{2}A_{01},&\nonumber \\
5)&{\cal M}(np\to \Sigma_c^0\overline{D^0} p)=
&-A_{11}&& +\sqrt{2}A_{01},&\nonumber \\
6)&{\cal M}(np\to \Sigma_c^+ D^- p)=&&A_{10} &-A_{01}, & \nonumber \\
7)&{\cal M}(np\to \Sigma_c^+\overline{D^0} n)=&&-A_{10} &-A_{01},& \nonumber \\
\hline
8)&{\cal M}(pp\to \Lambda_c^+\overline{D^0} p)=&A_{11},&& & \nonumber \\
9)&{\cal M}(np\to \Lambda_c^+{D^-} p)=
&\displaystyle\frac{1}{2}A_{11}&&&+\displaystyle\frac{1}{2}A_{00} , \nonumber \\
10)&{\cal M}(np\to \Lambda_c^+\overline{D^0} n)=
&\displaystyle\frac{1}{2}A_{11}&&&-\displaystyle\frac{1}{2}A_{00} ,  \nonumber \\
\end{array}
\end{equation} 
where $A_{I_1 I_2}$ are the isotopic amplitudes, corresponding to the total isospin $I_1$ for the initial nucleons and the total isospin $I_2$ for the produced $\overline{D}N$-system. In fact we have twice more of these reactions, by changing isospin $p\leftrightarrow n$ (for example 
$pp\to \Sigma_c^{++}{D^-} p\leftrightarrow nn\to \Sigma_c^0\overline{D^0} n$).

One can see, from Eq. (\ref{eq:amp}), that seven different matrix elements for different $N+N\to\Sigma_c+\overline{D}+N$-reactions, are characterized by three complex isotopic amplitudes: $A_{11}$, $A_{10}$ and $A_{01}$. Relations (\ref{eq:amp}) hold for any model in any kinematical condition.

Model independent relations among the seven differential cross sections hold:
\begin{equation}
\begin{array}{lrrr}
\sigma_1+ \sigma_2=&2|A_{11}|^2&+4|A_{10}|^2, & \nonumber \\
\sigma_3=&2|A_{11}|^2, & &\label{eq:sig} \\
\sigma_4+\sigma_5=
&2|A_{11}|^2& &+4|A_{01}|^2,\nonumber \\
\sigma_6+\sigma_7=
&&2|A_{10}|^2 &+2|A_{01}|^2.\nonumber 
\end{array}
\end{equation} 

The isotopic averaged cross section for the processes $N+N\to \Sigma_c+\overline{D}+N$, typically used in the transport codes, is defined as:
$$\overline{\sigma}=\displaystyle\frac{1}{4}\cdot 2\cdot\sum_{i=1-7}\sigma_i$$
Therefore:
$$\overline{\sigma}=3(|A_{11}|^2+|A_{10}|^2+|A_{01}|^2)$$
and does not contain any interference term between the different isotopic amplitudes.
The isotopic amplitudes can be expressed in terms of the cross sections as:
\begin{equation}
\begin{array}{ll}
|A_{11}|^2=&\displaystyle\frac{1}{2}\sigma_3,\nonumber \\
|A_{01}|^2=&\displaystyle\frac{1}{4}(\sigma_4+\sigma_5-\sigma_3) \label{eq:sam} \\
|A_{10}|^2=&\displaystyle\frac{1}{2}(\sigma_6+\sigma_7)-\displaystyle\frac{1}{4}(\sigma_4+\sigma_5-\sigma_3) \nonumber
\end{array}
\end{equation} 

The situation essentially changes near the reaction threshold, where all final particles are produced in $S$-state. The selection rules with respect to the Pauli principle, the P-invariance and the conservation of isospin and of the total angular momentum, allow to find a more simple  spin structure for the matrix element for $pp$ and $np$-collisions in  \cite{Re97}.
 A direct consequence of the threshold spin structure is that the interferences $A_{11}\bigotimes A_{01}^*$ and $A_{10}\bigotimes A_{01}^*$, which can be present in the general case, vanish in the threshold region, after summing over the polarizations of the colliding nucleons. Therefore at threshold we can derive the following  relations, additional to (\ref{eq:sig}),  which hold for any model: 
$$ \displaystyle\frac{d\sigma}{d\omega}(np\to \Sigma_c^{++}D^- n )= \displaystyle\frac{d\sigma}{d\omega}(np\to \Sigma_c^0\overline{D^0} p),$$ 
$$ \displaystyle\frac{d\sigma}{d\omega}(np\to \Sigma_c^+ D^- p)=\displaystyle\frac{d\sigma}{d\omega}(np\to \Sigma_c^+\overline{D^0} n).$$ 

\subsection{Scheme for estimation of cross sections}

Taking the following assumptions:
\begin{enumerate}
\item D-exchange model: the amplitudes are proportional to the amplitude for elastic DN-scattering, which is a combination of amplitudes for the two possible states of isotopic spin, ${\cal A}_1$ and $ {\cal A}_0$. 
\item SU(4) symmetry: 
${\cal A}_1\gg {\cal A}_0$ (as for KN scattering), so we neglect the amplitudes corresponding to $I(DN)=0$.
\item Threshold regime: all particles are in S final state; no interference between singlet and triplet initial states.
\end{enumerate}

We find:
$ \sigma_1=\sigma_2=\displaystyle\frac{1}{2}\sigma_3=|A_{11}|^2$, $\sigma_4=\sigma_5=|A_{11}|^2+2|A_{01}|^2=3\sigma_1$, and $\sigma_6=\sigma_7=|A_{01}|^2=\sigma_1$, 
$\sigma_9=\sigma_{10}=1/4\sigma_8$.

This derives from the fact that $A_{11}$ and $A_{01}$ are both proportional to ${\cal A}_1$, therefore the spin structure of theses amplitudes may be  different, but their moduli squared are the same. 

$\sigma_1$ can be calculated from the proportionality with $\sigma_8= \sigma(pp\to\Lambda_c^+\overline{D^0} p)$:
$\sigma_1=\displaystyle\frac{g^2_{p\Sigma_cD}}{g^2_{p\Lambda_cD}}\sigma_8$.

If one assume  $SU(4)$-symmetry, in the considered $D$-exchange model, the ratio of the coupling constants $\displaystyle\frac{g^2_{p\Sigma_cD}}{g^2_{p\Lambda_cD}} $ can be related to  $\displaystyle\frac{g^2_{p\Sigma K}}{g^2_{p\Lambda K}}$. An estimation of this ratio from the literature (\cite{Mo02} and refs herein) gives $\displaystyle\frac{g^2_{p\Sigma_cD}}{g^2_{p\Lambda_cD}}\simeq 0.1$. 

The overall $\overline{D^0}$-production, due to the discussed reactions, results in about $2\sigma_8$.

\subsection{Results}
 
The calculation \cite{Re01} takes into account the general threshold symmetry properties of the strong interaction, such as the the conservation of parity, total angular momentum, and the Pauli principle. 

Current models, which reproduce high energy data, can not always be safely extrapolated to threshold. For example, in \cite{Ca01} the cross section is taken as:
\begin{equation}
{\sigma_(pp\to\Lambda_c^+ \overline{D^0}p)}=a_X(1-Z)^{\alpha}Z^{-\beta}
\label{eq:eq3}
\end{equation}
with $a_X=0.496$, $Z=\sqrt{s^0_X}/\sqrt{s}$, $\sqrt{s^0_X}=5.069$ GeV, $\alpha=4.96$ and $\beta=1.36$ for $\overline{D^0}$ production, and slightly different paramaters for $D^-$-production:
$a_X=0.363$, $\sqrt{s^0_X}=5.073$ GeV, $\alpha=4.94$ and $\beta=1.44$.

\begin{figure}
\mbox{\epsfysize=15.cm\leavevmode \epsffile{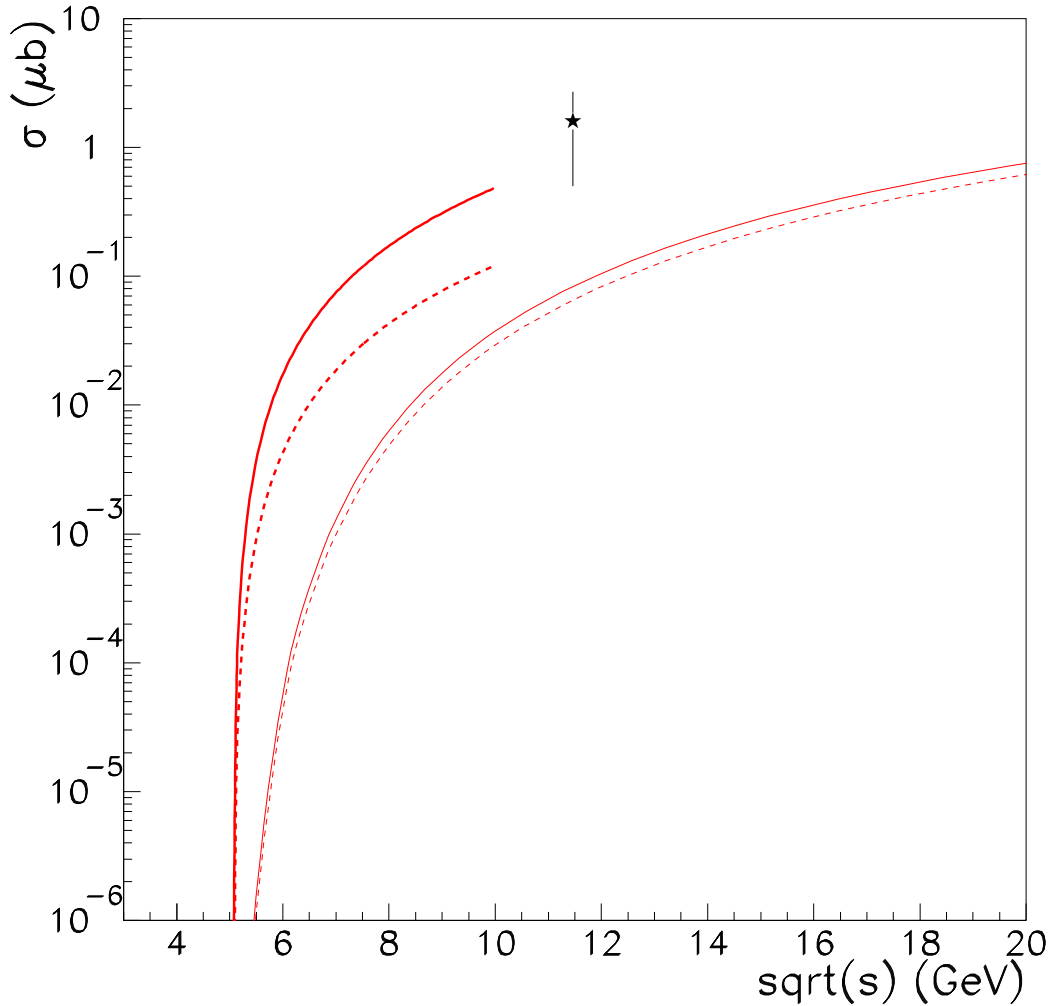}}
\vspace*{.2 truecm}
\caption{Comparison between the calculations \protect\cite{Re01} (thick lines)  and  \protect\cite{Ca01} (thin lines) for $\overline{D^0}$ production (solid lines) and for $D^-$-production (dashed lines).
}
\label{fig:fig3}
\end{figure}

In Fig. \ref{fig:fig3} a comparison between the calculations \cite{Re01} (thick lines)  and  \cite{Ca01} (thin lines) is shown for $\overline{D^0}$ production (solid lines) and for $D^-$-production (dashed lines).

One can see that the values can differ by an order of magnitude. This comes from the fact that the correct $Q$-behavior at threshold, from phase space considerations, should be quadratic, whereas the best fit parameters of Ref. \cite{Ca01} give an exponent $\simeq 5$. The difference is also due to the fact that the isotopic relations among $\overline{D^0}$ and $D^-$-cross sections are not respected in the parametrization \cite{Ca01}. This induces very large effect in particular in the threshold region.  

Note that the calculation \cite{Ga02}, which is based on the Quark-Gluon String model and the Regge phenomenology, is in agreement with \cite{Re01}, for a specific choice of form factors.

\subsection{Possible future studies}
 
One can foresee an extended program for the study of fundamental processes of charm hadroproduction, which will be useful for the experiments at the future international facility at GSI:
\begin{enumerate}
\item Analysis of vector charm $D^*$-associative production in $NN$-collisions
$N+N\to Y_c+D^*+N$;
\item Study of associative production of hperonic and nucleonic resonances:
$N+N\to Y_c^*+\overline{D}+N$ and $N+N\to Y_c+\overline{D}+N^*$;
\item Study of associative charm production in $\Delta N$, $\Delta \Delta $
and $BB$ collisions, where $BB$ is any baryonic or hyperonic resonance, which are important for high energy ion-ion collision;
\item Study of the transition regime for threshold to Regge description;
\item Consideration of diffractive production of charm particles in $pp$ and $p\overline{p}$-collisions.
\end{enumerate}

\section{Cross section for $p+p\to p+p+J/\psi$ at threshold}

A general formalism for the calculation of the cross section and the polarization observables for the reaction $p+p\to p+p+J/\psi$ at threshold has been developped in \cite{jpsi}. From a model independent parametrization of the matrix element, valid for all types of vector meson production in NN collisions, it is possible to derive expressions for the polarization phenomena and for the ratio of $J/\psi$ production in $np$ and $pp$ collisions, in terms of the threshold amplitudes. In framework of a model based on $t$-channel meson exchange and comparing with the existing data on $\phi$ production, it is possible to predict the threshold behavior of the cross section.

The ratio $R$ of cross sections for $\phi$ and $J/\psi $- production in framework of the same approach, namely for $\pi$-exchange in $N+N\to N+N+V^0$ and $\rho$-exchange for the subprocess $\pi+N\to N+V^0$, with $V^0=\phi$ or $J/\psi $, one can find:

$$R(J/\psi,\phi)=\displaystyle\frac{\sigma(pp\to ppJ/\psi)}{\sigma(pp\to pp\phi)}\simeq
\displaystyle\frac{g^2(J/\psi\to\pi\rho)}{g^2(\phi\to\pi\rho)}
\left (\displaystyle\frac{t_{\phi}-m_\pi^2}{t_{J/\psi}-m_\pi^2}\right )^2
\left (\displaystyle\frac{t_{\phi}-m_\rho^2}{t_{J/\psi}-m_\rho^2}\right )^2$$
$$
\cdot\displaystyle\frac{V_{ps}(J/\psi)}{V_{ps}(\phi)}
\displaystyle\frac{{\cal F}(\phi)}{{\cal F}(J/\psi)}
\left [\displaystyle\frac{F(t_{J/\psi})}{F(t_{\phi})}\right ]^2,$$
where $g(V\to \pi\rho)$ is the coupling constant for the decay $V\to \pi\rho$, $t_V=-mm_V$ is the threshold value of the momentum transfer squared, $F(t)$ is a phenomenological form factor for the vertex $\pi^*\rho^*V^0$, with virtual $\pi$ and $\rho$.

Taking the volume of three-particle phase space as \cite{Mo02}, one can find: 
$$\displaystyle\frac{V_{ps}(J/\psi)}{V_{ps}(\phi)}
=\displaystyle\frac{M_{J/\psi}}{M_{\phi}}
\left ( 
\displaystyle\frac{2m+M_{\phi}}{2m+M_{J/\psi}}\right )^{3/2}$$
and for the particle flux:
$$\displaystyle\frac{{\cal F}(\phi)}{{\cal F}(J/\psi)}=
\displaystyle\frac{k(\phi)}{k(J/\psi)}\left(
\displaystyle\frac{2m+M_{\phi}}{2m+M_{J/\psi}}\right ),$$
where $k$ is the momentum of the initial nucleon in the CM system:
$k=\sqrt{\displaystyle\frac{s}{4}-m^2}$, $s=(2m+m_V)^2$.

Using the existing experimental data about the decays 
$J/\psi\to \pi+\rho$ and $\phi\to \pi+\rho $ , one can find
$g^2(J/\psi\to\pi\rho)/g^2(\phi\to\pi\rho)\simeq 10^{-4}$, 

Taking the  terms:
$\left (\displaystyle\frac{t_{\phi}-m_\pi^2}{t_{J/\psi}-m_\pi^2}\right )^2=0.111$ GeV$^2$ and $\left (\displaystyle\frac{t_{\phi}-m_\rho^2}{t_{J/\psi}-m_\rho^2}\right )^2$=0.196, the ratio of phase space: 
$\displaystyle\frac{V_{ps}(J/\psi)}{V_{ps}(\phi)}\displaystyle\frac{{\cal F}(\phi)}{{\cal F}(J/\psi)}=0.216$, one finds for the ratio of cross sections, at the same $Q$:

$$\displaystyle\frac{\sigma(pp\to ppJ/\psi)}{\sigma(pp\to pp\phi)}
=4.7\cdot 10^{-7}$$.

The relevant available experimental data are: $\sigma(pp\to pp\phi)\simeq $ 300 nb at $p_L=3.67$ GeV \cite{Ba01}, and $\sigma_{exp}(pp\to pp J/\psi)=0.3\pm 0.09$ nb for $\sqrt{s}=6.7$ GeV (Q=1.725 GeV) \cite{Ba78}.

One can find  that
$$\sigma(pp\to pp\phi)\simeq 206\left ( \displaystyle\frac{Q}{0.1 \mbox{~GeV~}}\right ) ^2\mbox{~nb~}$$ 
and 

\begin{equation}
\sigma(pp\to pp J/\psi)\simeq  9.7\cdot 10^{-5}\left ( \displaystyle\frac{Q}{0.1 \mbox{~GeV~}}\right ) ^2 \left [F(t_{J/\psi})/F(t_{\phi})\right ]^2 \mbox{~nb~}. 
\label{jpsi}
\end{equation}

This value is too small, when compared with the existing experimental value for the lowest $\sqrt{s}=6.7$ GeV, namely $\sigma_{exp}(pp\to ppJ/\psi)=0.3\pm 0.09$ nb.

Note that the $\rho$-exchange model for $\sigma(\pi N\to J/\psi)$ gives a cross section one order of magnitude smaller in comparison with other possible theoretical approaches \cite{Bo75,Ko79,Be76}. It is one possibility to explain the value of $\sigma_{exp}(pp\to pp J/\psi)$. Another possibility is to take
$\left [F(t_{J/\psi})/F(t_{\phi})\right ]^2\simeq 10$, which can be plausible, because the $J/\psi=c\overline{c}$-system must have a smaller size in comparison with $\phi=s\overline{s}$. This can be realized by the following form factor:
$$F_V(t)= \displaystyle\frac{1}{1-\displaystyle\frac{t}{\Lambda_V^2}},$$
with ${\Lambda_V}\simeq m_V$.

The cross section, based on Eq. (\ref{jpsi}) normalized the the experimental point at $\sqrt{s}=6.7$ GeV, i.e., taking the ratio 
$\left [F(t_{J/\psi})/F(t_{\phi})\right ]^2=10$, is plotted in Fig. \ref{jpsi}, together with the experimental data from the compilation \cite{Vogt}, where different symbols differentiate $J\psi$ production in $pp$ or extrapolated from $pA$ collisions. Note, in conclusions, that in the framework of the considered model, one can find:
$$\displaystyle\frac{\sigma(np\to np J/\psi)}{\sigma(pp\to pp J/\psi)}=5,$$
which would require a correction of the experimental data on $pA$ reaction, where equal $np$ and $pp$ cross sections are usually assumed for the extraction of the elementary cross section.
\begin{figure}
\mbox{\epsfysize=15.cm\leavevmode \epsffile{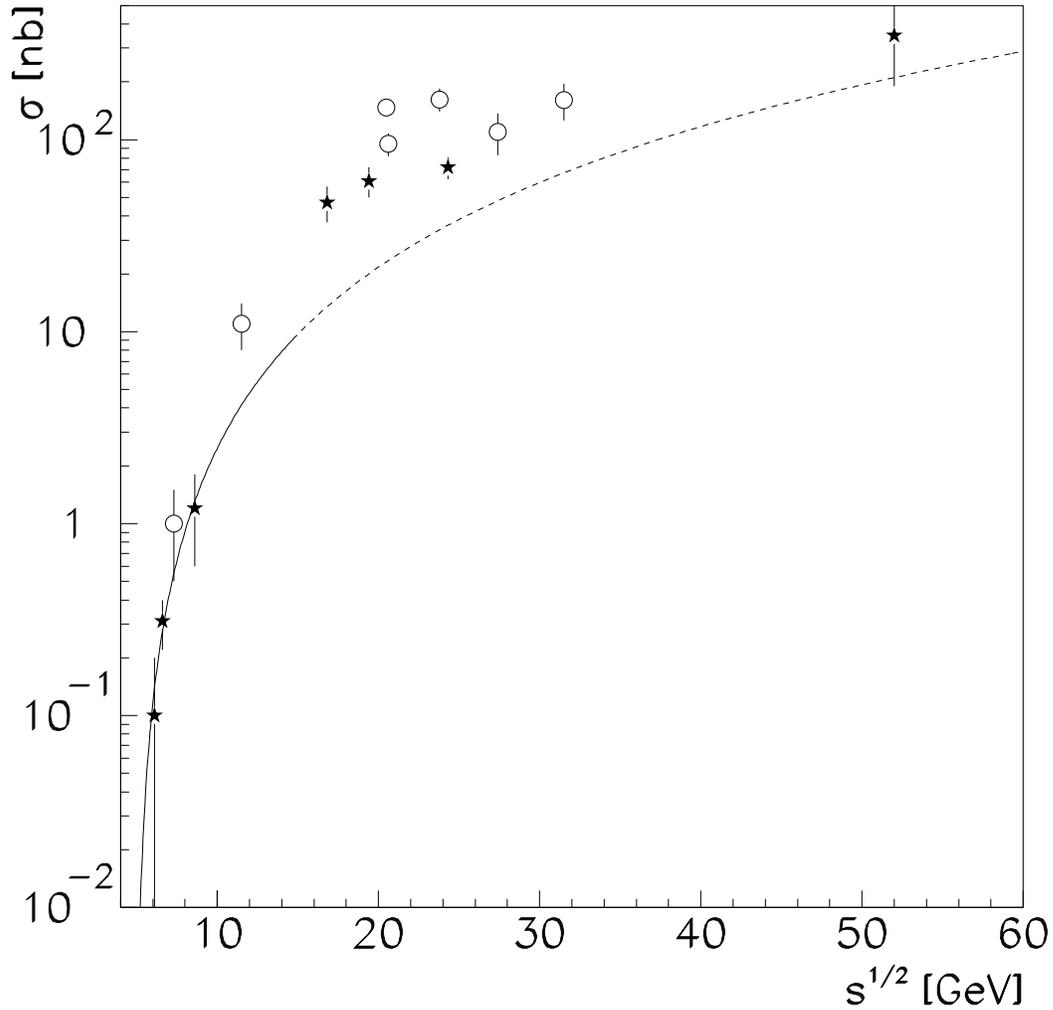}}
\vspace*{.2 truecm}
\caption{ Cross section for $J/\psi$ production in $pp$ collisions. Data are from \protect\cite{Vogt}
}
\label{fig:jpsi}
\end{figure}

{}

\end{document}